\documentclass[conference]{IEEEtran}
\IEEEoverridecommandlockouts
\usepackage{cite}
\usepackage{amsmath,amssymb,amsfonts}
\usepackage{algorithmic}
\usepackage{graphicx}
\usepackage{textcomp}
\usepackage{booktabs}

\usepackage{xcolor}
\def\BibTeX{{\rm B\kern-.05em{\sc i\kern-.025em b}\kern-.08em
    T\kern-.1667em\lower.7ex\hbox{E}\kern-.125emX}}
\begin{document}

\title{Multi-programming Cross Platform Benchmarking for Quantum Computing Hardware
	%TODO: add microsoft for funding
\thanks{This work is funded by the QuantUM Initiative of the Region Occitanie, University of Montpellier and IBM Montpellier. The access to Quantinuum machine is funded by Microsoft Azure.}
}

\author{\IEEEauthorblockN{Siyuan Niu}
	\IEEEauthorblockA{\textit{LIRMM, University of Montpellier} \\
		161 Rue Ada, Montpellier, France \\
		siyuan.niu@lirmm.fr}
	\and
	\IEEEauthorblockN{Aida Todri-Sanial}
	\IEEEauthorblockA{\textit{LIRMM, University of Montpellier, CNRS} \\
		161 Rue Ada, Montpellier, France \\
		aida.todri@lirmm.fr}
	
}

\maketitle

\begin{abstract}
With the rapid development of quantum hardware technologies, benchmarking the performance of quantum computers has become attractive. In this paper, we propose a new aspect of benchmarking quantum computers by evaluating the limitation of hardware utilization using a multi-programming mechanism -- a technique that simultaneously executes multiple circuits in a quantum machine. This is the first attempt to compare the evaluation of multi-programming on trapped-ion and superconducting devices. Based on the experimental results, performing multi-programming on a trapped-ion device demonstrates better results than a superconducting machine without losing any fidelity to independent executions.
\end{abstract}


\section{Introduction}
Quantum computing is a rapidly growing research field in recent years. Various quantum platforms have been developed based on different technologies, for example, superconducting, trapped-ion, photonics, neutral-atom, etc. The two leading technologies are superconducting and trapped-ion devices. Users can access the real quantum machines through online services released by companies such as Google, IBM, or Quantinuum.

The near-term quantum machines with at most around 100 qubits are qualified as Noisy Intermediate-Scale Quantum (NISQ) hardware~\cite{preskill2018quantum}. The hardware topology is limited and the quantum operations are prone to errors. Even with these hardware constraints, the fast improvement of quantum hardware leads to two frequently asked questions: (1) Which metric to use to characterize the performance of a quantum computer? (2) How to use a quantum computer more efficiently given the currently available hardware resources while obtaining reliable results? 

To address the first question, quantum volume (QV)~\cite{cross2019validating} is one of the most used metrics to benchmark NISQ devices proposed by IBM. The best QV that IBM quantum machine obtained is 128, whereas the trapped-ion device produced by Quantinuum was announced to achieve 4096 for QV. Other single-number metrics such as Q-Score~\cite{martiel2021benchmarking} or application-based benchmarks such as SupermarQ~\cite{tomesh_supermarq_2022} have also been developed. As for the second question, the multi-programming mechanism was introduced for superconducting and DWAVE machines by enabling parallel circuits or QUBO executions on one quantum chip simultaneously~\cite{das2019case,pelofske2022parallel}. The multi-programming mechanism can help improve hardware utilization, reduce the overall problem-solving time, and even save costs for charging quantum machines.

In our paper, we combine the two questions by benchmarking the hardware ability to enable multi-programming mechanism across two different platforms: a superconducting device from IBM and a trapped-ion device from Quantinuum. First, we compare the characteristics of the two machines. Second, we evaluate the multi-programming mechanism and report the fidelity difference compared with the standalone circuit execution mode.

\section{Comparison of Quantum Hardware Characteristics}
\begin{table}[!htp]
	\caption{Characteristics of the two quantum devices.}
	\centering

		\begin{tabular}{c c c}
			\toprule                              
			Vendors & IBM & Quantinuum\\
			\midrule
			Backend name& ibmq\_mumbai& H1-2 \\
			\midrule
			Technology& Superconducting &Trapped-ion \\
			\midrule
			Qubits & 27 & 12\\
			\midrule
			QV & 128 & 4096\\
			\midrule
			Gate sets & $Rz, SX, X, CX$& $U_{1q}, Rz, ZZ$\\
			\midrule
			1Q Error (\%) & 0.02 & 0.01 \\
			\midrule
			2Q Error (\%) & 4.5 & 0.35 \\
			\midrule
			RO Error (\%) & 2.9 & 0.4 \\
			\midrule
			Topology & Nearest-neighbor & All-to-all \\
			\midrule
			Parallel two-qubit operation & All connected qubits & 3 parallel zones\\
		
			\bottomrule
		\end{tabular}

	\label{tab:devices}
\end{table}

We compare ibmq\_mumbai, a 27-qubit IBM quantum machine with the largest QV, with a 12-qubit System Model H1-2 Quantinuum device. Their characteristics are shown in Table~\ref{tab:devices}. H1-2 machine demonstrates more reliable quantum operations compared to ibmq\_mumbai, the 1Q, 2Q, and readout error being improved by 2x, 12.9x, and 7.2x. Moreover, the QCCD architecture~\cite{pino2021demonstration} used by the H1-2 machine enables multiple interaction zones so that the parallel two-qubit operations become possible.

\section{Multi-programming Mechanism Benchmarking}
The multi-programming mechanism was first proposed to apply to the superconducting device because of its relatively large number of qubits. 
The fidelities of circuits are decreased due to a limited number of reliable qubits and the higher probability of crosstalk. Even though the size of the current trapped ion device is relatively small, it is still interesting to investigate if the same issues of crosstalk and non-uniform reliable qubits occur in the trapped-ion device. Additionally, since all of the online services for trapped-ion machines are charged per circuit, executing multiple circuits in one execution can save budget as well. In this section, we first execute two circuits on both devices in parallel to report the fidelity difference between simultaneous and standalone executions. Second, we simultaneously apply QAOA twice to evaluate its performance in solving the Max-Cut problem using multi-programming mechanism.

\subsection{Simultaneous Executions of Two Circuits}
\begin{table}[!htp]
	\caption{Information of benchmarks.}
	\centering
	
	\begin{tabular}{c c c c c}
		\toprule                              
		ID & Benchmark & Qubits & Gates & CX\\
		\midrule
		1&BV3 & 3& 9& 2\\
		\midrule
		2&BV4 & 4 & 12&3 \\
		\midrule
		3&peres\_3 & 3 &16 &7\\
		\midrule
		4&toffoli & 3 & 15&6\\
		\midrule
		5&3\_17\_13 & 3& 36&17\\
		\midrule
		6&4mod5-v1\_22 & 5 & 21& 11\\
		\midrule
		7&mod5milds\_65 & 5 & 35&16 \\
		\midrule
		8&alu-v0\_27 & 5 & 36&17 \\
		\midrule
		9&decod24-v2\_43 & 4 & 52&22 \\
		
		\bottomrule
	\end{tabular}

	\label{tab:benchmarks}
\end{table}

The benchmarks used to evaluate the performance of multi-programming are collected from the state of the art~\cite{niu2021enabling,das2019case}, shown in Table~\ref{tab:benchmarks}. For ibmq\_mumbai, we apply QuMC~\cite{niu2021enabling} multi-programming algorithm and set the optimization level of qiskit transpiler to the highest level of three. The version of Qiskit is 0.25.0 and the number of shots is set to 8192. Whereas for Quantinuum H1-2, the transpilation process is not accessible by users and we only set the number of shots to 100 to maintain a reasonable budget. Moreover, we use the metric Probability of Successful Trials (PST), defined as the number of trials giving the correct answer divided by the total number of trials, to represent the fidelity of the circuit. 

\begin{figure}[!h]
	\centering
	
		\includegraphics[scale=0.5]{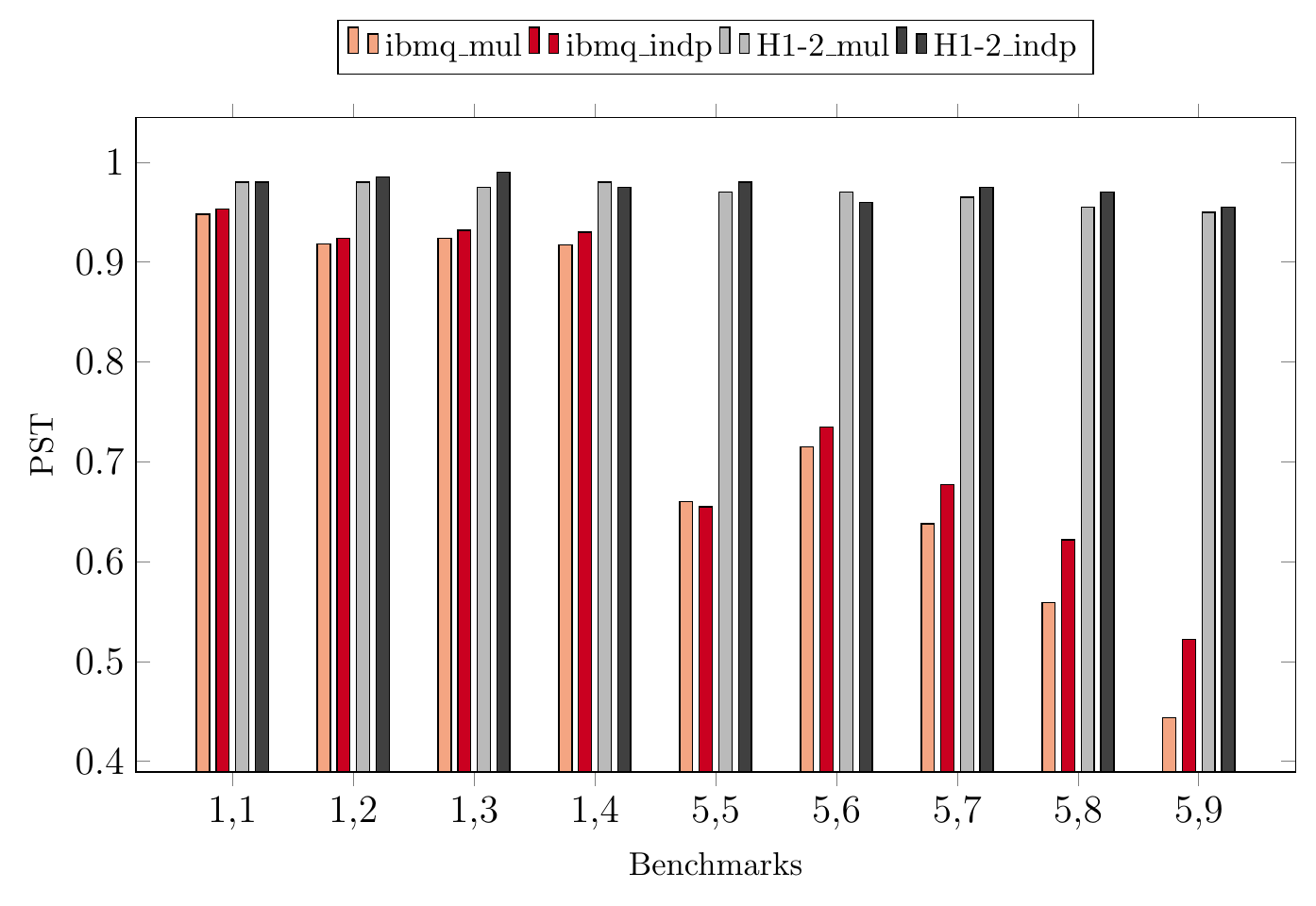}
	
	\caption{Results of executing two benchmarks simultaneously and independently on ibmq\_mumbai and Quantinuum H1-2. 
	}
	\label{fig:pst}
	
\end{figure}

The results of executing two benchmarks simultaneously or independently are shown in Fig.~\ref{fig:pst}. In general, Quantinuum H1-2 machine obtains more reliable results than ibmq\_mumbai. The simultaneous executions on ibmq\_mumbai (labeled as ibmq\_mul) decrease the fidelities by 3.4\% compared to standalone executions (labeled as ibmq\_indp). Whereas for Quantinuum H1-2, similar results are obtained comparing simultaneous (labeled as H1-2\_mul) with independent executions (labeled as H1-2\_indp), with only a difference of 0.5\% on fidelities, but the budget of simultaneous executions is reduced by 31\%. Therefore, enabling multi-programming on Quantinuum trapped-ion devices can improve the hardware utilization, reduce the total circuit runtime (execution time + waiting time), and save the budget without losing fidelity, which also demonstrates the low impact of crosstalk.

\subsection{Simultaneous Executions of QAOA}
Inspired by the parallel executions of multiple QUBO problems on DWAVE machine~\cite{pelofske2022parallel}, we perform QAOA algorithm twice on the two machines at the same time and evaluate its performance in solving Max-Cut problem. 

For demonstration, we construct an QAOA ansatz for a 4-node rectangle graph and optimize the parameters using \emph{COBYLA} optimizer. We execute two of the same circuits with optimized parameters on the two quantum chips. For Quantinuum H1-2, if we execute the QAOA ansatz once independently, the probabilities of obtaining the correct answers ``0101'' and ``1010'' are the highest in the probability distribution, and the sum of the two probabilities is 50\%. When executing two of the same ansatz in parallel, both of the ansatz circuits are able to have the highest probabilities for the correct answers, the sum being 64\% and 47\%, respectively. Also, the budget is reduced by 30\%. Whereas for ibmq\_mumbai, executing QAOA ansatz once independently and twice simultaneously obtain correct answers with highest probabilities as well, the sum being 42.5\% (independent), 38.1\%, and 40.3\% (the last two for simultaneous executions).

\section{Conclusion}
Cross platform benchmarking becomes an active research topic with the rapid improvement of different quantum technologies. Various metrics have been proposed to characterize the performance of a quantum computer. We propose a novel aspect of evaluating the hardware limitation by performing multi-programming mechanism, a technique that enables to execute multiple circuits in parallel to improve the hardware utilization and reduce the total circuit runtime. We first execute two small benchmarks on IBM superconducting quantum chip and Quantinuum trapped-ion machine. Second, we run two QAOA ansatz circuits in parallel on the two machines to solve the Max-Cut problem. Based on the results, we found that trapped-ion devices are more suitable for multi-programming mechanism, without losing fidelities compared with independent executions, and the cost budget is reduced significantly.

%\section*{Acknowledgment}
%
%The preferred spelling of the word ``acknowledgment'' in America is without 
%an ``e'' after the ``g''. Avoid the stilted expression ``one of us (R. B. 
%G.) thanks $\ldots$''. Instead, try ``R. B. G. thanks$\ldots$''. Put sponsor 
%acknowledgments in the unnumbered footnote on the first page.

%
\bibliographystyle{plain}

\bibliography{bibliography}{}

\end{document}